\documentstyle[twoside,fleqn,espcrc2,epsfig]{article}

\newcommand{\Tr}{{\mathrm{Tr}}}
\newcommand{\mod}{{\mathrm{mod}}}
\newcommand{\dual}{\mbox{}^{\ast}}
\newcommand{\dd}{\mbox{\rm d}}

\newcommand{\Z}{{Z \!\!\! Z}}
\newcommand{\beqn}{\begin{eqnarray}}
\newcommand{\eeqn}{\end{eqnarray}}
\newcommand{\eq}[1]{(\ref{#1})}
\newcommand{\cD}{{\cal D}}
\newcommand{\cC}{{\cal C}}
\newcommand{\cZ}{{\cal Z}}
\newcommand{\LL}{{I\!\!  L}}
\newcommand{\CK}[1]{\mbox{\scriptsize c}_{\mbox{$\scriptstyle #1$}}}
\newcommand{\nddsum}[2]{\sum_{\stackrel{\scriptstyle \dual
#1(\dual\CK{#2}) \in \Z} {\delta \dual #1=0}}}

\def\bbbone{{\mathchoice {\rm 1\mskip-4mu l} {\rm 1\mskip-4mu l}
{\rm 1\mskip-4.5mu l} {\rm 1\mskip-5mu l}}}
\def\bbbc{{\mathchoice {\setbox0=\hbox{$\displaystyle\rm C$}\hbox{\hbox
to0pt{\kern0.4\wd0\vrule height0.9\ht0\hss}\box0}}
{\setbox0=\hbox{$\textstyle\rm C$}\hbox{\hbox
to0pt{\kern0.4\wd0\vrule height0.9\ht0\hss}\box0}}
{\setbox0=\hbox{$\scriptstyle\rm C$}\hbox{\hbox
to0pt{\kern0.4\wd0\vrule height0.9\ht0\hss}\box0}}
{\setbox0=\hbox{$\scriptscriptstyle\rm C$}\hbox{\hbox
to0pt{\kern0.4\wd0\vrule height0.9\ht0\hss}\box0}}}}
\def\bbbe{{\mathchoice {\setbox0=\hbox{\smalletextfont e}\hbox{\raise
0.1\ht0\hbox to0pt{\kern0.4\wd0\vrule width0.3pt
height0.7\ht0\hss}\box0}}
{\setbox0=\hbox{\smalletextfont e}\hbox{\raise
0.1\ht0\hbox to0pt{\kern0.4\wd0\vrule width0.3pt
height0.7\ht0\hss}\box0}}
{\setbox0=\hbox{\smallescriptfont e}\hbox{\raise
0.1\ht0\hbox to0pt{\kern0.5\wd0\vrule width0.2pt
height0.7\ht0\hss}\box0}}
{\setbox0=\hbox{\smallescriptscriptfont e}\hbox{\raise
0.1\ht0\hbox to0pt{\kern0.4\wd0\vrule width0.2pt
height0.7\ht0\hss}\box0}}}}

\title{
~\vspace{-1.5cm}
\begin{flushright}
{\large ITEP-TH-46/98}
\end{flushright}
\vspace{0.5cm}
Aharonov--Bohm effect, Center Monopoles and Center Vortices in
$SU(2)$ Lattice Gluodynamics}

\author{M.N.~Chernodub, M.I.~Polikarpov, A.I.~Veselov
and M.A.~Zubkov
\address{Institute of Theoretical and Experimental Physics,
B.Cheremushkinskaya 25, Moscow, 117259, Russia}}

\begin{document}

\begin{abstract}
$SU(2)$ gluodynamics is investigated numerically and
analytically in the (Indirect) Maximal Center gauge at finite temperature.
The center vortices are shown to be condensed in the confinement
phase and dilute in the deconfinement phase. A new physical object, center
monopole, is constructed. We show that the center monopole condensate is
the order parameter of deconfinement phase transition.
The linking of the vortex worldsheets and quark trajectories is identified
with the Aharonov--Bohm interaction in an effective Abelian Higgs theory.
We conclude that the confinement in the Maximal Center gauge can be
explained by a new mechanism called "{\it the \underline{real}
superconductor mechanism"}.  
\end{abstract}

\maketitle

\section{Introduction}

The investigation of the confinement
in $SU(N)$ gauge theories is mainly based on a partial gauge fixing of
the nonabelian gauge group up to its abelian subgroup. The popular
abelian gauge, the so called "Maximal Center gauge", is
proposed in Ref.~\cite{CenterGaugeFirst}. In this gauge
the nonabelian gauge group is fixed up to its center subgroup. In a
center gauge the $SU(N)$ gauge theory is reduced to a $\Z_N$ gauge theory
which contains vortex strings as topological defects. Lattice
calculations~\cite{NumerousCenter} in the Maximal Center
gauge show that the dynamics of these defects plays an important role
in the color confinement. Below we study the central vortices and new
topological defects, "center monopoles": and we discuss confinement
mechanism in the Maximal Center gauge.

\section{Center Vortices and Center Monopoles in the Maximal Center
Projection}

We study $SU(2)$ gluodynamics with the standard Wilson action.
The Maximal Center gauge makes the link matrices $U$ as close to the
center elements of $SU(2)$ group ($\pm \bbbone$) as possible.
This gauge is defined as follows~\cite{CenterGaugeFirst}:
first we fix the Maximal Abelian gauge by maximizing the field
functional~\cite{MaA}, $\sum\nolimits_{l} \Tr (U_l \sigma^3 U^+_l
\sigma^3)$, over gauge transformations
$U^{(\Omega)}_{x,\mu} = \Omega^+_x
U_{x,\mu} \Omega_{x+\hat\mu}$, the summation is
taken over links $l$ of the lattice,
$\sigma^a$ are the Pauli matrices. Second we maximize the functional
$\sum\nolimits_{l} \cos^2 \arg({(U_l)}^{11} )$
over residual
$U(1)$ gauge transformations, this
fixing makes the link matrices close to the central elements $\pm
\bbbone$. 

The center vortices are defined as follows~\cite{CenterGaugeFirst}. After
fixing the Maximal Center gauge we define the $\Z_2$ plaquette variables
$\sigma_P$:
\beqn
\sigma_P \equiv {(\dd n)}_P = n_1 + n_2 - n_3 - n_4\,,
\label{sigma}
\eeqn
where links $1,\dots,4$ form the boundary of the plaquette $P$
and $n_l = {\mathrm{sign}} ({\mathrm{Tr}} U_l)$. The
worldsheet of the center vortex is defined on the dual lattice as a
collection $\dual \sigma$ of the plaquettes dual to the non-zero
plaquettes $\sigma_P$, the worldsheets $\dual \sigma$ are closed on the
dual lattice ($\delta \dual \sigma = 0$).

The interaction of the center vortices with the Wilson loop is
topological. To see this we represent the $SU(2)$ gauge
field $U_l$ in the Maximal Center gauge as a product of the $\Z_2$
variable $\exp\{i \pi n_l\}$, $n_l=0,1$, and the $SU(2) \slash \Z_2$
variable $V_l$, $\Tr V_l \ge 0$: $U_l = \exp\{i \pi n_l \} \cdot
V_l$. Taking into account \eq{sigma} we rewrite the Wilson loop for
the contour $\cC$ as:
\beqn
W_\cC = \Tr \prod\limits_{l \in C} U_l =
\exp\Bigl\{i \pi \LL (\cC,\sigma) \Bigr\}
\Tr \prod\limits_{l \in C} V_l\,,
\label{WC1}
\eeqn
where $\LL(\cC,\sigma)$ is the linking number of the quark trajectory
$\cC$ with the string worldsheet~$\dual \sigma$~\cite{PoWiZu}:
\beqn
\LL (\cC,\sigma) = (\sigma,m[\cC]) = (\sigma,\Delta^{-1} \dd \cC)\,,
\label{LL}
\eeqn
$m[\cC]$ is a surface spanned on the contour $\cC$: $\delta m[\cC] = \cC$.
The last expression for $\LL$ is the four-dimensional analogue of the
Gauss formula describing the linking of the closed surface $\dual \sigma$
and closed loop $\cC$. 

In is known~\cite{CenterGaugeFirst,LinkingConf} that the topological
interaction $\exp\{i \pi \LL\}$ gives a correct value for the string
tension. Below we study the properties of $\Z_2$ excitations in the
Maximal Center projection.

Due to $\Z_2$ periodicity the theory contains monopole-like excitations
("center monopoles"). The monopole trajectories are defined as follows:
\beqn
j = \frac{1}{2} \dd \Bigl[(\dd n) \, \mod \, 2 \Bigr]
\equiv \frac{1}{2} \dd \Bigl[\sigma \, \mod \, 2 \Bigr]\,.
\eeqn
The $\Z_2$ charge of the center mo\-no\-po\-le is con\-ser\-ved, the
monopole tra\-jec\-to\-ri\-es are closed:~$\delta \dual j = 0$.
\begin{figure}[!htb]
\begin{minipage}{7.5cm}
\begin{center}
\begin{tabular}{cc}
\epsfig{file=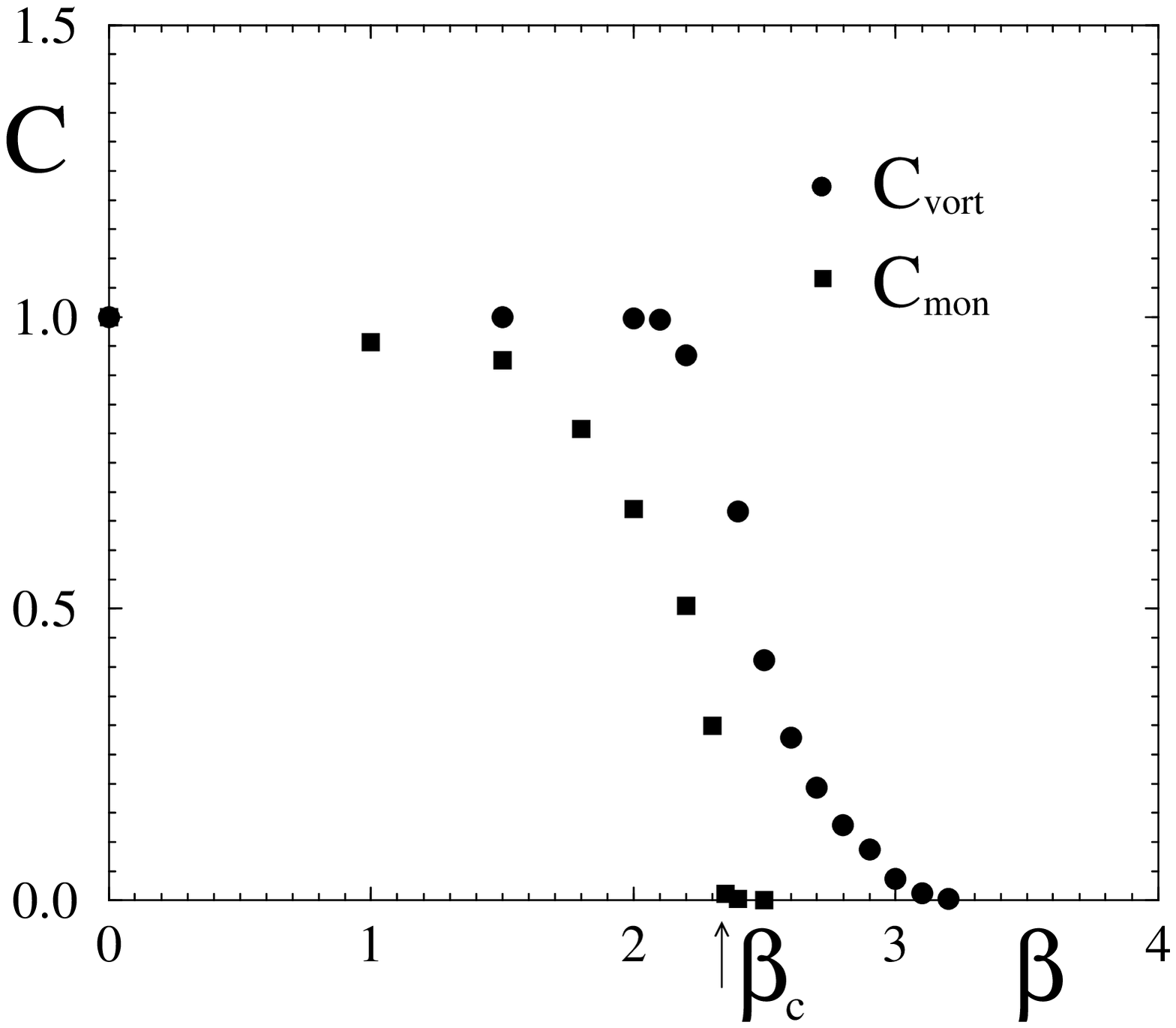,width=3.2cm,height=3.5cm,angle=0}&
\epsfig{file=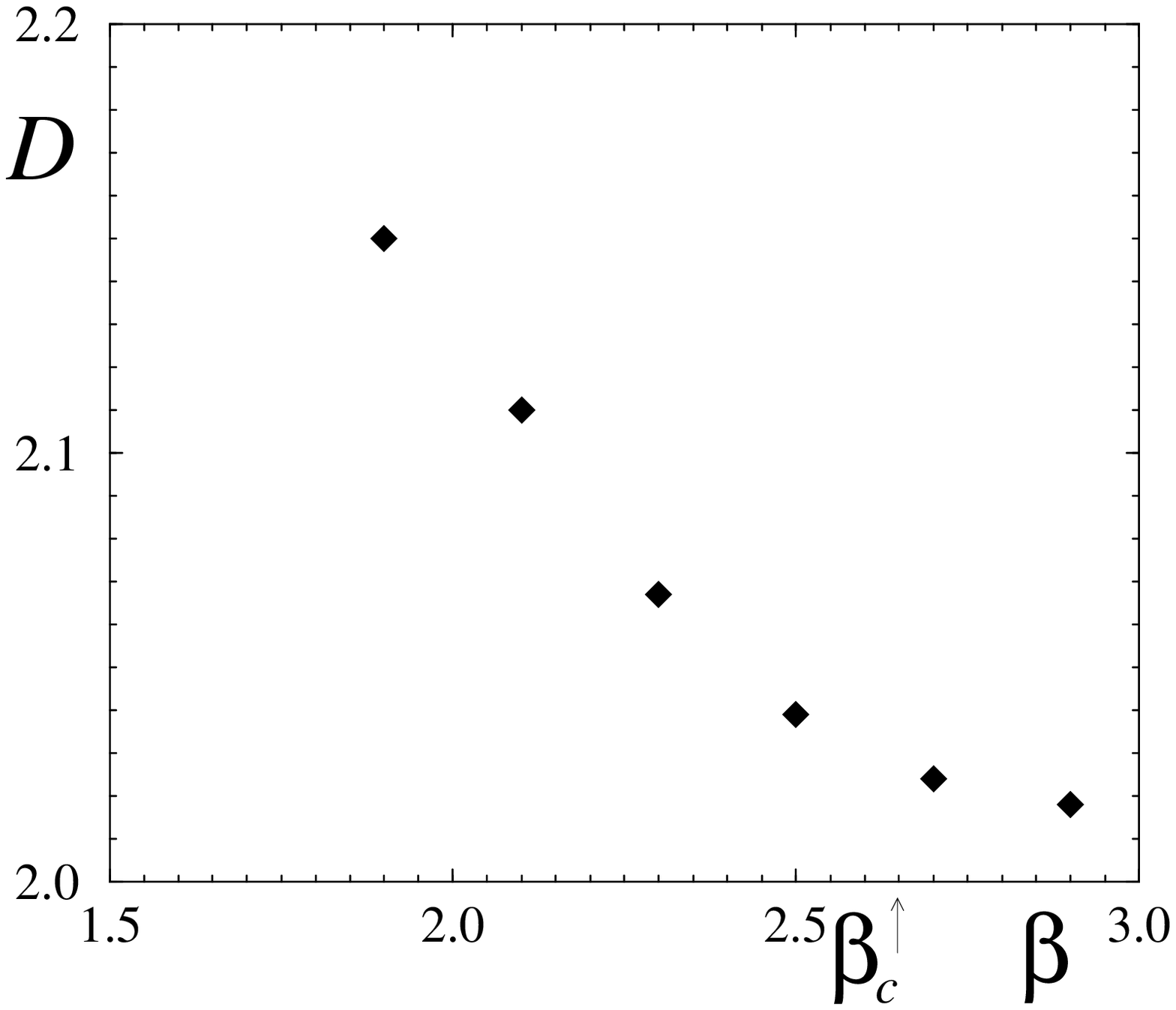,width=3.2cm,height=3.5cm,angle=0}\\
(a) & (b)\\
\end{tabular}
\vspace{-7mm}
\caption[]{\small (a) Percolation probability for center vortex strings
$C_{\mathrm{vort}}$ and center monopoles $C_{\mathrm{mon}}$ {\it vs.}
$\beta$ on $16^3 \times 4$ lattice; (b) Fractal dimension $D$ of the
vortex string network on $12^3 \times 8$ lattice.}
\end{center}
\end{minipage}
\end{figure}
\vspace{-5mm}

The important dynamical property of the center monopoles is the 
percolation probability $C_{\mathrm{mon}}$ which is defined as a 
probability for two different points of the lattice to be connected 
by {\it the same} center monopole trajectory~\cite{Percolation}. 
We observe, that $C_{\mathrm{mon}}$ vanishes in the 
deconfinement phase, and is non-zero in the confinement phase
($C_{\mathrm{mon}}$ is 
shown by boxes in Figure~1(a)). We conclude that the confinement phase 
transition is accompanied by the condensation of the center 
monopoles. The monopoles are dual abelian degrees of freedom and 
their condensation means that the confinement phase corresponds to 
the dual superconductor phase for the 
Maximal Center gauge~\cite{InPrep}.  On the other hand, in the
next  Section we show that the confinement in the Maximal Center gauge 
might be explained by a different mechanism.

\section{Real Superconductor Mechanism in the Maximal Center gauge}

The partition function of the Maximal Center gauge is:
\beqn
\cZ_{g.f} = \int \cD U \, e^{- \beta \sum\nolimits_P (1 -
\frac{1}{2} \Tr U_P)  - S_{g.f.}[U]}\,,
\label{Zgf}
\eeqn
where the action $S_{g.f.}$ includes the Faddeev--Popov determinant and
gauge fixing functionals.

The action $S_{g.f.}[U]$ is invariant under the
trans\-for\-ma\-tions $U \to - U$, therefore $S_{g.f.}[U] = S_{g.f.}[V]$, 
where
$U = e^{i \pi n} \, V$, $\Tr V > 0$, $n=0,1$. Using this property the
quantum average of the Wilson loop can be represented as follows:
\beqn
<W_\cC>  = \frac{1}{\cZ_{g.f.}} \, \int\limits_{\Tr V > 0} \cD V \,
e^{- S(V)} \, \Tr \prod\limits_{l \in C} V_l \nonumber\\
\cdot \nddsum{\sigma}{2} e^{ - \beta \sum\limits_P \Tr V_P \cdot (1 -
\cos(\pi \sigma_P))  + i \pi \LL (\cC,\sigma) }\,,
\label{WC2}\\
S(V) = \beta \sum\limits_P (1 - \frac{1}{2} \Tr V_P )
+ S_{g.f.}[V]\,.\nonumber
\eeqn
To derive eq.\eq{WC2} we used the definition of the vortex strings
\eq{sigma} and the representation for the Wilson loop \eq{WC1}.

The interaction proportional to linking number of the world sheet $\dual 
\sigma$ and test particle world trajectory $\cC$ is already known in the
field theory~\cite{Alford,PoWiZu}. This is the Aharonov--Bohm (AB)
interaction of the text particle which scatters on the string carrying a
magnetic flux. Below we show how to rewrite the considered theory in terms
of the Abelian Higgs theory. The world sheets of 
the Abrikosov--Nielsen--Olesen (ANO)
vortex strings~\cite{Abrikosov} in this theory corresponds to variables 
$\dual \sigma$ (center vortices).

The expectation value for the Wilson loop~\eq{WC2} can be represented as:
\beqn
<W_\cC>  = \frac{1}{\cZ_{g.f.}} \, \int\limits_{\Tr V > 0} 
\cD V \, e^{- S(V)} \, \Tr \prod\limits_{l \in C} V_l \nonumber\\
\lim\limits_{\kappa \to + \infty} \int\limits^{+ \pi}_{ - \pi} \!\!  \cD
\theta \!\! \int\limits^{+\pi}_{-\pi} \!\!  \cD \varphi \, 
e^{ - {\tilde S}(\theta,\varphi;V) + i (\theta,\cC)}\,, \label{Z1}
\eeqn
where the effective abelian action is:
\beqn
{\tilde S}(\theta,\varphi;V) = \beta \sum\limits_P \Tr V_P
\cdot (1 - \cos{(\dd \theta)}_P) \nonumber\\ + \kappa \sum\limits_l (1 -
\cos(\dd \varphi + 2 \theta))\,.
\label{A1}
\eeqn
Here $\theta$ is the compact abelian gauge field, $\varphi$ is the phase
of the Higgs field $\Phi = |\Phi|\, e^{i \varphi}$ and the radial part of
the Higgs field in the effective abelian theory is
frozen (this corresponds to the London limit of the theory). The Higgs
field carries the double charge. The couplings in the effective abelian
theory fluctuate due to external integration over the field~$V$.

To prove that \eq{Z1} is equivalent to \eq{WC2} we have to fix the unitary
gauge, $\dd \varphi = 0$, and note that in the limit $\kappa \to \infty$
the integral over the variable $\theta$ is reduced to the sum over $\pi
n$, $n=0,1$. The ANO vortex strings are defined as $\sigma = \dd n$. 

The mechanism of confinement in the AHM representation \eq{Z1} is as
follows. The confinement phase of gluodynamics corresponds to the Coulomb
phase of the effective AHM \eq{Z1} since in the Coulomb phase the ANO
vortices are condensed.  The abelian Higgs field carries the electric
charge $e=2$, therefore the vortices carry the magnetic flux $2 \pi \slash
e = \pi$ Since the ANO strings carry non-trivial flux they interact with
the electrically charged particles (quarks) via AB effect: the
interaction is proportional to the linking number of the vortex
worldsheets with the world trajectory of the charged particle \eq{LL}.
According to numerical calculations~\cite{CenterGaugeFirst,LinkingConf}
this topological interaction reproduces the $SU(2)$ string tension. We
call the described mechanism of confinement as "{\it real superconductor
mechanism}" since in this picture the electrically charged particles are
to be condensed in order to provide the suppression of the vortex strings
in the deconfinement phase. 

One may expect that the the AB interaction is strong in the confinement
phase since the network of the ANO strings is percolating.  The
percolation probability for vortices $C_{\mathrm{vort}}$ is defined
similarly to that for center monopoles (see previous Section). We show the
quantity $C_{\mathrm{vort}}$ {\it vs.} $\beta$ on $16^3\times 4$ lattice
in Figure~1(a) by circles.  It is clearly seen that in the confinement
phase the vortex strings are percolating with the maximal
probability $C_{\mathrm{vort}}=1$.  The quantity
$1-C_{\mathrm{vort}}$ is an order parameter for the phase transition. 

In Figure~1(b) we show the fractal dimension of the vortex string
network, $D = 1+2A/L$ on $12^3 \times 8$ lattice Here $A$ is the number
of plaquettes and $L$ is the number of links on the string. The fractal
dimension $D$ is high at the confinement phase what is a characteristic
feature of a percolating vortex network. In the deconfinement phase the
value of this quantity is close to $2$ since we have the dilute string
ensemble.

\section*{Acknowledgments}

The authors are grateful to J.~Greensite and E.T.~Tomboulis for useful
discussions. This work was supported by the grants INTAS-96-370,
INTAS-RFBR-95-0681, RFBR-96-02-17230a and RFBR-96-15-96740.


\begin{thebibliography}{99}
\bibitem{CenterGaugeFirst} L.Del~Debbio {\it et al.},
Phys.Rev. D55 (1997) 2298.
\bibitem{NumerousCenter} L.Del~Debbio {\it et al.}, hep-lat/9802003.
\bibitem{MaA} A.S.~Kronfeld, M.L. Laursen, G. Schierholz, U.J.
Wiese, Phys.Lett. 198B (1987) 516.
\bibitem{PoWiZu} M.I.~Polikarpov, U.J.~Wie\-se and M.A.~Zub\-kov,
{\it Phys.Lett.} {\bf B309} (1993) 133.
\bibitem{Alford} M.G.~Alford and F.Wilczek, {Phys.Rev.Lett.}, {62}
(1989) 1071; M.G.~Alford, J.~March--Russel and F.Wilczek, {
Nucl.Phys.}, { B337} (1990) 695; J.~Preskill and L.M.~Krauss, {
Nucl.Phys.}, { B341} (1990) 50.
\bibitem{LinkingConf} L.Del~Debbio {\it et~al.}, 
Nucl.Phys.Proc.Suppl. 63 (1998) 552.
\bibitem{Percolation} A.V.~Pochinsky, M.I.~Polikarpov and
B.N. Yur\-chen\-ko, Phys.Lett.~A154 (1991) 194;\\
T.L.~Ivanenko, A.V.~Pochinskii and M.I.~Po\-li\-kar\-pov,
Phys.Lett.~B302 (1993) 458.
\bibitem{InPrep} M.N. Chernodub, M.I. Po\-li\-kar\-pov, A.I. Ve\-se\-lov
and M.A. Zub\-kov, in preparation.
\bibitem{Abrikosov} A.A. Abrikosov, Sov.Phys. JETP 32 (1957) 1442;
H.B. Nielsen and P.Olesen, Nucl.Phys. B61 (1973) 45.
\end{thebibliography}
\end{document}